\documentclass[twocolumn,aps,prd,nobibnotes,byrevtex,nofootinbib,superscriptaddress,amsfonts,amsmath,amssymb,floatfix,10pts]{revtex4-1}
\usepackage{hyperref}
\usepackage{enumerate}
\hypersetup{
    colorlinks=true,
    citecolor=blue,
    linkcolor=blue,
    filecolor=magenta,      
    urlcolor=blue,
}

\usepackage{charter}
\usepackage{subfig}
\usepackage{graphicx}
\usepackage{multirow}
\usepackage{dcolumn}
\usepackage{mathrsfs}
\usepackage{acronym}
\usepackage[usenames,dvipsnames,svgnames]{xcolor}
\usepackage{soul}
\usepackage{tabularx}

\newcommand{\orderAoneB}{\ensuremath{10^{-13}}}
\newcommand{\orderAtwoB}{\ensuremath{10^{-26}}}
\newcommand{\orderBoneB}{\ensuremath{10^{-14}}}
\newcommand{\orderBtwoB}{\ensuremath{10^{-27}}}
\newcommand{\aprAoneB}{15.3}
\newcommand{\aprAtwoB}{-33.0}
\newcommand{\aprBoneB}{10.2}
\newcommand{\aprBtwoB}{4.17}
\newcommand{\msAoneB}{8.92}
\newcommand{\msAtwoB}{-4.78}
\newcommand{\msBoneB}{4.87}
\newcommand{\msBtwoB}{-21.2}
\newcommand{\mpaAoneB}{12.0}
\newcommand{\mpaAtwoB}{-16.7}
\newcommand{\mpaBoneB}{12.0}
\newcommand{\mpaBtwoB}{-60.7}
\newcommand{\wffAoneB}{16.8}
\newcommand{\wffAtwoB}{-42.3}
\newcommand{\wffBoneB}{17.4}
\newcommand{\wffBtwoB}{-46.1}
\newcommand{\omegaGR}{\textcolor{black}{\ensuremath{1.68 \times 10^{-9}}}}
\newcommand{\ratioGR}{\textcolor{black}{\ensuremath{10^{-3}}}}
\newcommand{\wGR}{\textcolor{black}{\ensuremath{1.68 \times 10^{-10}}}}
\newcommand{\rGR}{\textcolor{black}{\ensuremath{10^{-2}}}}
\newcommand{\msun}{\ensuremath{M_{\odot}}}
\newcommand{\kret}{\ensuremath{\mathscr{K}}}

\begin{document}
\title{Neutron star structure in Ho\v{r}ava-Lifshitz gravity}
\author{Kyungmin Kim}
\email[]{kkim@kasi.re.kr}
\affiliation{Korea Astronomy and Space Science Institute, Daejeon 34055, Republic of Korea}

\author{John J. Oh}
\email[]{johnoh@nims.re.kr}
\affiliation{National Institute for Mathematical Sciences, Daejeon 34047, Republic of Korea}

\author{Chan Park}
\email[]{iamparkchan@gmail.com}
\affiliation{National Institute for Mathematical Sciences, Daejeon 34047, Republic of Korea}

\author{Edwin J. Son}
\email[]{eddy@nims.re.kr}
\affiliation{National Institute for Mathematical Sciences, Daejeon 34047, Republic of Korea}

\date{\today}

\begin{abstract}
We present interesting aspects of neutron stars (NSs) from the standpoint of a modified theory of gravity called Ho\v{r}ava-Lifshitz (HL) gravity. A deviation from general relativity (GR) in HL gravity  can change typical features of the NS structure. In this study, we investigate the NS structure by deriving the Tolman-Oppenheimer-Volkoff equation in HL gravity. We find that a NS in HL gravity with a larger radius and heavier mass than a NS in GR remains stable without collapsing into a black hole.
\end{abstract}

\maketitle

%% Acronyms
\acrodef{hl}[HL]{Ho\v{r}ava-Lifshitz}
\acrodef{ks}[KS]{Kehagias-Sfetsos}
\acrodef{tov}[TOV]{Tolman-Oppenheimer-Volkoff}
% \acrodef{dbc}[DBC]{detailed balance condition}
\acrodef{gr}[GR]{General relativity}
\acrodef{uv}[UV]{ultraviolet}
\acrodef{ir}[IR]{infrared}
\acrodef{sbh}[SBH]{Schwarzschild black hole}
\acrodef{ksbh}[KSBH]{KS black hole}
\acrodef{EOS}[EOS]{equation-of-state}
\acrodef{ns}[NS]{neutron star}
\acrodef{nss}[NSs]{neutron stars}
\acrodef{apr4}[APR4]{Akmal-Pandharipande-Ravenhall}
\acrodef{ms1}[MS1]{M{\"u}ller-Serot}
\acrodef{mpa1}[MPA1]{M{\"u}ther-Prakash-Ainsworth}
\acrodef{wff1}[WFF1]{Wiringa-Fiks-Fabrocini}
\acrodef{sly}[SLy]{Skyrme-Lyon}
\acrodef{gw}[GW]{gravitational wave}
\acresetall

%%%%%%%%%%%
%%  Introduction  %%
%%%%%%%%%%%
\section{Introduction}
\label{sec:intro}
\ac{gr} has been a successful theory of gravity for explaining the motion of planets and stars at the macroscopic scale up to now, because it is strongly valid under the weak gravitational field approximation. However, the theory is believed to be theoretically incomplete in terms of a unified theory with a quantum nature, for example, in describing the physics at strong gravitational fields near compact astrophysical objects such as black holes or \acp{ns}, mysterious \emph{darkness} such as dark energy/matter, the origin of the Universe such as inflationary blowup, and the hierarchical phase transition at the very early stage of the Universe (see~ Refs. \cite{Freese:2017idy,Silk:2013opt,Tsujikawa:2013fba,Guth:1980zm,Olive:1989nu,Brandenberger:2000as,Boyanovsky:2006bf} and the references therein). Although there have been many attempts to address these problems by introducing new particles and/or alternative theories of gravitation, a breakthrough is still required from the theoretical as well as experimental/observational point of view.

One of the discrepancies between gauge theories and \ac{gr} is that the latter is not renormalizable due to \ac{uv} divergence. The \ac{uv} divergence appears in extreme circumstances and/or at high energies, where the quantum effect has to be taken into account, for example, at the early stage of the Universe or in the vicinity of a black hole. A renormalizable gravity theory is naturally supposed to have improved properties at the \ac{uv} scale. One of the theories embodying such a philosophy has been proposed by Ho\v{r}ava~\cite{Horava:2008ih,Horava:2009uw,Horava:2009if} with the aim of formulating \iac{uv}-complete theory by sacrificing local Lorentz symmetry, which is called \ac{hl} gravity. \ac{hl} gravity approximates to \ac{gr} at the \ac{ir} scale, whereas it becomes a different type of gravity at the \ac{uv} scale by the introduction of anisotropic scaling between time and space. This anisotropic scaling is key to renormalizability in \ac{hl} gravity although it breaks the local Lorentz symmetry. Many intensive studies in diverse fields of applications have been conducted in search of a possible candidate for quantum gravity (for recent progress reports, see Refs.~\cite{Blas:2018flu, Wang:2017brl, Mukohyama:2010xz} and the references therein).

Further, \emph{the deformed \ac{hl} gravity} has been introduced to obtain asymptotically flat solutions by introducing a parameter $\omega$ in addition to the parameters in the original \ac{hl} gravity~\cite{Horava:2009uw,Park:2009zra}. A static spherically symmetric and asymptotically flat solution has been found by Kehagias and Sfetsos (\acsu{ks})~\cite{Kehagias:2009is} that approaches the \ac{sbh} at the \ac{ir} limit.\footnote{Birkoff's theorem in \ac{hl} gravity has been studied in Ref.~\cite{Devecioglu:2018ote}, which states that the theorem is still valid for solutions allowing the \ac{gr} limit in the \ac{ir} region, whereas it is violated for the unusual solutions in the \ac{uv} region. The \ac{ks} solution we considered here is the unique case of the GR limit admitting Birkhoff’s theorem. Therefore, we have chosen the \ac{ks} solution as an exterior space of a neutron star.} Among several parameters in the deformed \ac{hl} gravity, $\omega$ is the only remaining free parameter appearing in the \ac{ks} solution. The \ac{ks} solution in the deformed \ac{hl} gravity has been constrained by several observations in Refs.~\cite{Liu:2010hzb,Horvath:2011xr,Iorio:2010cs}. By introducing a dimensionless form $\tilde{\omega} = \omega (G_N c^{-2} M)^2$, the parameter is bounded from below, $\tilde{\omega}_\mathrm{min} \sim 10^{-17}$, $10^{-16}$, and $10^{-15}$, from light deflections by Jupiter, Earth, and the Sun, respectively~\cite{Liu:2010hzb}. $\tilde{\omega}_\mathrm{min} \sim 10^{-16}$ also originates from the weak lensing by galaxies of mass $\sim 10^{10} \msun$~\cite{Horvath:2011xr}. In addition, $\tilde{\omega}_\mathrm{min} \sim 10^{-10}$ is derived from the orbital motion of the S2 star~\cite{s2star:2009aa} and the range residual of Mercury~\cite{Iorio:2010cs}. These lower bounds are converted to $\omega_\text{min} \sim 10^{-48}-10^{-16} \ \text{cm}^{-2}$.

Here, one might ask how extreme circumstances can affect the structural equilibrium between \ac{hl} gravity and isotropic matter. Tolman~\cite{Tolman:1939jz} and Oppenheimer and Volkoff \cite{Oppenheimer:1939ne} (\acsu{tov})~formulated the so-called \acsu{tov} equation in \ac{gr} to describe the equilibrium state of compact stars such as white dwarfs and \acp{ns} in a spherically symmetric and static configuration. By solving the \ac{tov} equation, we can obtain the mass-radius relation and, consequently, can estimate the maximum mass of a compact star and its radius. In particular, for a \ac{ns}, the maximum mass and radius are sensitive to the \emph{stiffness} of the selected \ac{EOS} model. Hence, by investigating the TOV equation with the realistic \ac{EOS} parameters, we can precisely understand the UV aspect of a \ac{ns} in \ac{hl} gravity in comparison with the case in \ac{gr}. There have been many studies with diverse approaches to the stellar structure in \ac{hl} gravity, for example, the stellar magnetic field configuration and the solution of Maxwell's equations in \ac{hl} gravity as well as in the modified $f(R)$ gravity \cite{Hakimov:2013zoa}, black holes and stars with the projectability condition \cite{Greenwald:2009kp}, and gravitational collapse in \ac{hl} gravity \cite{Greenwald:2013kja}. Other attempts at spontaneous scalarization in scalar-tensor gravity for compact stars have been studied in Refs.~\cite{PhysRevD.58.124003, PhysRevD.99.124022, Rosca-Mead:2020bzt}. Various aspects of a \ac{ns} in different modified gravities were intensively studied in Refs.~\cite{Astashenok:2013vza, Astashenok:2014pua, Astashenok:2014gda, Astashenok:2014nua, Kim:2013nna, Harko:2013wka, Prasetyo:2017hrb, Barausse:2011pu}.

In this study, we investigate the structure of a \ac{ns} in \ac{hl} gravity, in comparison with that in \ac{gr}. 
We derive an explicit form of the \ac{tov} equations in \ac{hl} gravity. By solving them with realistic \ac{EOS} models, we can obtain the mass-radius relation of a \ac{ns}, which estimates how \ac{hl} gravity deviates from \ac{gr} in terms of the mass and the radius of a \ac{ns}.
In Sec.~\ref{sec:HL}, the \ac{tov} equation of perfect fluids in \ac{hl} gravity is derived for this purpose. We then solve the equation in Sec.~\ref{sec:NS} with the selected \ac{EOS} models of \ac{apr4}~\cite{APR:1998}, \ac{mpa1}~\cite{MPA1:1987}, \ac{ms1}~\cite{MS1:1975}, and \ac{wff1}~\cite{WFF:1988} because, from the \ac{tov} calculation in \ac{gr}, these models have shown results that are consistent with the observed maximum mass, $\sim$$2\msun$ of \ac{ns}s~\cite{2solar:2010nature,2solar:2018apj}. For the numerical computation of the \ac{tov} equation, we adopt the fifth-order Runge-Kutta solver~\cite{DORMAND198019,Shampine:1986:SPR}. The mass-radius relation in \ac{hl} gravity is compared with that in \ac{gr}, which shows that both the mass and the radius of the heaviest \ac{ns} in  \ac{hl} gravity are larger than those in \ac{gr}. It is shown that the distance-dependent profile of the pressure decreases more gradually, and the surface condition where the pressure becomes zero is satisfied at farther radii due to the relatively weaker gravitational attraction of \ac{hl} gravity than that of \ac{gr}. We address the detailed investigation into the mass-radius relation in \ac{hl} gravity as well as the parametric limit of the \ac{hl} parameter $\omega$ based on the result.
Finally, we discuss our results in terms of observational validation in Sec.~\ref{sec:discussion}.

%%%%%%%%%%%%%%%%%%%%%%%%%%%%%%%%%%
\section{TOV equation in HL gravity}
\label{sec:HL}
The action of \ac{hl} gravity is formulated by an anisotropic scaling between time and space, $t \rightarrow b^{z}  t$ and $x^i \rightarrow b x^i$ and the form of $z=3$:  
\begin{eqnarray}
I_\text{\acs{hl}} &=& \int dt d^3x \sqrt{g} N \left[\frac{2}{\kappa^2} \left( {\mathcal K}_{ij}{\mathcal K}^{ij} - \lambda {\mathcal K}^2 \right) \right.\nonumber\\ 
&&-\left. \frac{\kappa^2}{2\zeta^4} \left( {\mathcal C}_{ij} - \frac{\mu\zeta^2}{2} {\mathcal R}_{ij} \right) \left( {\mathcal C}^{ij} - \frac{\mu\zeta^2}{2} {\mathcal R}^{ij} \right)\right.\nonumber\\
&&+ \! \left. \! \frac{\kappa^2\mu^2(4\lambda \! - \! 1)}{32(3\lambda \! - \! 1)} \! \! \left( \! {\mathcal R}^2 \! + \! \frac{4( \omega \! - \! \Lambda_W )}{4\lambda \! - \! 1} {\mathcal R} \! + \! \frac{12 \Lambda_W^2}{4\lambda \! - \! 1} \! \right) \! \right]
\label{action}
\end{eqnarray}
with the \emph{softly} broken detailed balance condition,
which is sometimes called the deformed \ac{hl} gravity in the literature. Note that ${\mathcal K}_{ij} \equiv \frac{1}{2N} [ \dot{g}_{ij} - \nabla_i N_j - \nabla_j N_i ]$ is an extrinsic curvature, where $N$ is a lapse function, $N^{i}$ is a shift vector, $g_{ij}$ is a three-dimensional spatial metric, and $\dot{g}_{ij}$ denotes $\partial{}g_{ij}/\partial{}t$. ${\mathcal C}^{ij} \equiv \varepsilon^{ik\ell} \nabla_k ( {\mathcal R}_\ell^j -  \delta_\ell^j {\mathcal R}/4 )$ is a Cotton-York tensor, where ${\mathcal R}_{ij}$ and ${\mathcal R}$ are a three-dimensional spatial Ricci tensor and a Ricci scalar, respectively. $\kappa^2$ is a coupling related to the Newton constant $G_N$, and $\lambda$ is an additional dimensionless coupling constant. Here, $\omega$ is essential for an asymptotically flat solution, though it violates the detailed balance condition. The coupling constants $\mu$, $\Lambda_W$, and $\zeta$ stem from the three-dimensional Euclidean topologically massive gravity action~\cite{Deser:1981wh, Deser:1982vy}. When $\lambda=1$, the Einstein-Hilbert action can be recovered in the \ac{ir} limit by identifying the fundamental constants with $c = (\kappa^2 / 4) [\mu^2 (\omega-\Lambda_W) / (3\lambda-1)]^{1/2}$, $G_N = \kappa^2 c^2 / 32 \pi$, and $\Lambda = -(3 / 2) \Lambda_W^2 / (\omega-\Lambda_W)$, representing the speed of light, the gravitational constant, and the cosmological constant, respectively.\footnote{It has been claimed that the original model proposed by Ho\v{r}ava has no nontrivial solution for the nonprojectable case through constraint analysis~\cite{Henneaux:2009zb}. However, the study is restricted to the case of nontrivial coupling constant $\lambda \ne 1$. Because we consider only $\lambda=1$ for reproducing \ac{gr} in the \ac{ir} regime (in addition, $\Lambda=0$ for simplicity), the inconsistency of the original \ac{hl} gravity claimed above can be avoided in our paper. See Ref.~\cite{Devecioglu:2020dny} for more analyses of constraint dynamics in arbitrary dimensions.}

The total action under consideration is given by $I_\text{tot} = I_\text{\acs{hl}} + I_\text{mat}$, where $I_\text{mat}$ represents the matter action that will be specified by assuming a perfect fluid and choosing an \ac{EOS} without an explicit form. In addition, we consider hereafter an asymptotically flat geometry, that is, $\Lambda = 0$, for simplicity.

Now, if we consider a static, spherically symmetric metric ansatz,
\begin{equation}
\label{eq:metric}
ds^2 = - e^{2\Phi(r)} c^2 dt^2 \! + \! \frac{dr^2}{1 \! - \! f(r)} \! + \! r^2 \! \left( d\theta^2 \! + \! \sin^2\theta d\phi^2 \right),
\end{equation}
then the equations of motion in \ac{hl} gravity with the stress-energy tensor of a perfect fluid, $T_{\mu\nu} = (\rho+p) u_{\mu}u_{\nu} + pg_{\mu\nu}$, are given by
\begin{subequations}
\label{eom}
\begin{align}
% \frac{\kappa}{4} \rho c^2 &= \frac{\kappa^4 \mu^2}{64 r^2} \left( 2 r \omega f + \frac{f^2}{r} \right)', \label{eom:rho} \\
\rho &= \frac{c^2}{16\pi G_{N} r^2 \omega} \left( 2 r \omega f + \frac{f^2}{r} \right)', \label{eom:rho} \\
% 8 \pi p &= \frac{\kappa^4 \mu^2}{64 r^4} \left[ f \left( f - 2 r^2 \omega \right) + 4 r \left( 1 - f \right) \left( f + r^2 \omega \right) \Phi' \right], \label{eom:p} \\
p &= \frac{c^4}{16\pi G_{N} r^4 \omega} \left[ f \left( f - 2 r^2 \omega \right) + 4 r \left( 1 - f \right) \left( f + r^2 \omega \right) \Phi' \right], \label{eom:p} \\
p' &= - \left( \rho c^2 + p \right) \Phi', \label{eom:cons}
\end{align}
\end{subequations}
where $u_{\mu}=(1,0,0,0)$ is a four-vector field, $\rho$ and $p$ are the energy density and the pressure of a perfect fluid, respectively, and the prime denotes $d/dr$. Note that \ac{hl} gravity has six parameters, as observed in Eq.~\eqref{action}: we have already fixed $\lambda$ and $\Lambda_W$ to 1 and 0, respectively, $\kappa$ and $\mu$ are hidden in the physical constants related to $c$ and $G_N$ in the \ac{ir} region, and $\zeta$ does not contribute to this nonrotating configuration. Thus, $\omega$ is the only remaining free parameter of the theory under consideration.

To solve Eq.~\eqref{eom}, we replace $f(r)$ by $m(r)$ through the relation $f = -r^2\omega + \sqrt{r \omega ( r^3 \omega + 4 G_N c^{-2} m )}$. Then, when $m(r) = M$, it approaches $f\xrightarrow{\omega\to\infty} 2 G_N c^{-2} M / r - 2 G_N^2 c^{-4} \omega^{-1} M^2 / r^4 + \cdots$, where the first term is nothing but the Schwarzschild solution, and the following terms are \ac{hl} corrections depending upon the $\omega$ parameter. Here, the mass parameter $M$ is identified as the quasilocal energy at infinity~\cite{Myung_2010}. For the \ac{ks} solution, we see $e^{2\Phi(r)} = 1 - f(r)$, while Eq.~\eqref{eom:cons} governs the behavior of the  $(tt)$-component of the metric, which is coupled to matter in the generic case. Note that the horizons of the \ac{ks} black hole are located at $r_\pm = G_N c^{-2} \left( M \pm \sqrt{M^2 - M_c^2}\right)$, where $M_c\equiv (2 \omega G_N^2 c^{-4})^{-1/2}$ is the critical mass of the horizons formed. With the Planck mass $m_P$ and length $l_P$, $M_c$ can be rewritten as $M_c = m_P / \sqrt{2 \omega \ell_P^2}$, which reduces to $M_c = m_P / \sqrt{2}$ for $\omega = \ell_P^{-2}$ and becomes $M_c \approx \msun$ for $\omega \approx 2.293 \times 10^{-11} \, \text{cm}^{-2}$. For a constant $\omega$, the black hole horizon exists with $M\ge M_{c}$.\footnote{When $M < M_{c}$, a naked singularity appears, and the detailed behavior of the formation of naked singularities and wormholes in Ho\v{r}ava gravity is studied in~\cite{Bellorin:2014qca,Son:2010bs}. In the quantum regime, however, the \ac{ks} geometry turns out to be regular~\cite{Gurtug:2017kqf}, which may safely avoid the naked singularity problem.} On the other hand, for a fixed $M$, one finds $\omega \ge \omega_{c} \equiv (2 G_N^2 c^{-4} M^2)^{-1}$ to form a black hole. Hence, the horizon of a solar mass \ac{ks} black hole exists only if $\omega \ge \omega_{c}\approx 2.293 \times 10^{-11} \textrm{cm}^{-2}$. Requiring that a $4 \msun$ black hole candidate, GRO J0422+32~\cite{Gelino:2003pr}, has the horizon, we have $\omega \gtrsim 1.433 \times 10^{-12} \textrm{cm}^{-2}$. In this manner, we restrict the range of the parameter as $\omega \ge 2 \times 10^{-12} \ \textrm{cm}^{-2}$.

Now, we can rewrite Eq.~\eqref{eom} as follows:
\begin{subequations}
\label{eq_tov}
\begin{align}
m' &= 4 \pi r^2 \rho, \label{Eq_tov_mass} \\
p' &= \frac{(\rho c^2 + p) r \omega \left[ \left( 1 + \tilde{\rho} \right) - \sqrt{1 + 4 \tilde{\rho}} - \tilde{p} \right]}{\sqrt{1 + 4 \tilde{\rho}} \left[ 1 + r^2\omega \left( 1 - \sqrt{1 + 4 \tilde{\rho}} \right) \right]}, \label{eq_tov_pre}
\end{align}
\end{subequations}
where $\tilde{\rho} = G_N c^{-2} \omega^{-1} m r^{-3}$ and $\tilde{p} = 4 \pi G_N c^{-4} \omega^{-1} p$. Note that if we expand $p'$ in terms of $1/\omega\rightarrow 0$ using the relation $\sqrt{1+\epsilon} \approx 1 + \epsilon/2 + \cdots$ for $\epsilon \ll 1$ and take the leading order, we obtain
\begin{equation}
 p' \approx - \frac{G_N m \rho}{r^2} \left(1 + \frac{p}{\rho c^2}\right) \left( 1 + \frac{4 \pi r^3 p}{m c^2 } \right) \left(1 - \frac{2 G_N m}{r c^{2}}\right)^{-1},
\end{equation}
which reproduces the \ac{tov} equation in \ac{gr}.

%%%%%%%%%%%%%%%%%%%%%%%%%%%%%%%%%%
\section{NS Structure in HL gravity} 
\label{sec:NS}
For the study of the \ac{ns} structure in \ac{gr}, various \ac{EOS} models have been taken into account. Among them, we, in particular, select four \ac{EOS} models that result in the maximum mass, $M_\mathrm{max}$, of a \ac{ns} becoming similar to or heavier than $2\msun$ from the conventional \ac{tov} calculation in \ac{gr}, and the results agree with the recent observations of $\sim$$2\msun$ \ac{ns}~\cite{2solar:2010nature,2solar:2018apj}. The selected \ac{EOS} models are \ac{apr4}~\cite{APR:1998}, \ac{mpa1}~\cite{MPA1:1987}, \ac{ms1}~\cite{MS1:1975}, and \ac{wff1}~\cite{WFF:1988}: \ac{mpa1} and \ac{ms1} are, respectively, derived using the relativistic Brueckner-Hartree-Fock theory and the relativistic mean field theory. \ac{apr4} and \ac{wff1} are derived with the variational method but with different nucleon potential models (see Ref.~\cite{APR:1998} for the difference in the used models). All of these models assume that the nuclear matter contains neutrons, protons, electrons, and muons. In addition, to imitate the crust structure near the surface of \iac{ns}, we replace the low-density region of the considered \ac{EOS} models with the Skyrme-Lyon model~\cite{SLy:2001} as studied in~\cite{Read:2008iy}.

To solve $m$ and $p$ in Eq.~\eqref{eq_tov} simultaneously, we adopt the fifth-order Runge-Kutta method which controls the error with the fourth-order method~\cite{DORMAND198019,Shampine:1986:SPR}. For the computation, we first set $\rho\left(0\right)=\rho_{c}$, $p\left(0\right)=p_{c}$ at the center $r=0$, where we call $\rho_c$ and $p_c$ as the central density and pressure, respectively. We obtain $m\left(r\right)$ and $p\left(r\right)$ at each $r$ by taking $\left(\rho_{c},p_{c}\right)$ from each of the \acp{EOS} and solving Eq.~\eqref{eq_tov} numerically. The radius $R$ is then determined by $R = r^*$ where $p(r=r^*) = 0$, and the mass is simultaneously determined by $M = m(R)$. Thus, by varying $\rho_{c}$, we produce the series of $\left(M,R\right)$ called the mass-radius relation, which describes the static profile of the spherically symmetric \ac{ns} structure for a given \ac{EOS} model. The exterior of the \ac{ns} is naturally described by the \ac{ks} vacuum solution $e^{2\Phi(r > R)} = 1 - f(r > R) = 1 + r^2\omega - \sqrt{r \omega ( r^3 \omega + 4 G_N c^{-2} M )}$, and an integration constant of the metric function $\Phi(r)$ is fixed by this boundary condition at $r=R$.

%%  Mass-Radius plots: start  %%
\begin{figure}[tbp]
  \centering
  \subfloat[\label{fig:mrplot_a} 
  \ac{apr4} and \ac{ms1} case
  ]{\includegraphics[width=\columnwidth]{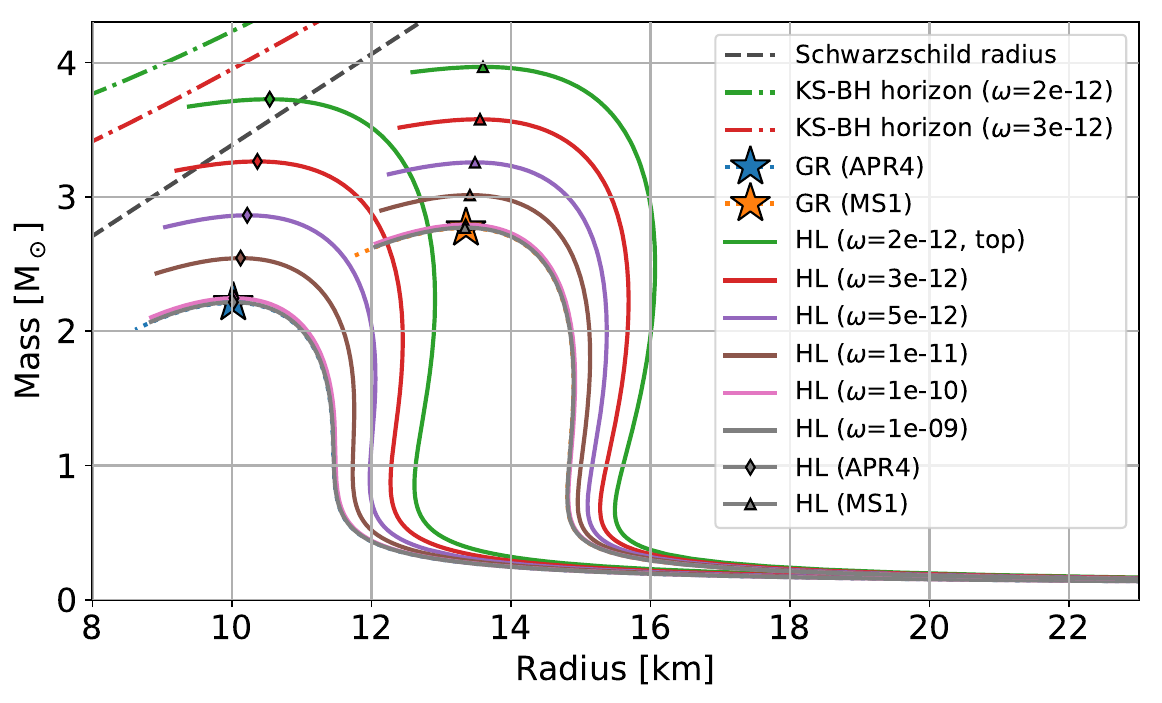}}\\
  \subfloat[\label{fig:mrplot_b} 
  \ac{mpa1} and \ac{wff1} case
  ]{\includegraphics[width=\columnwidth]{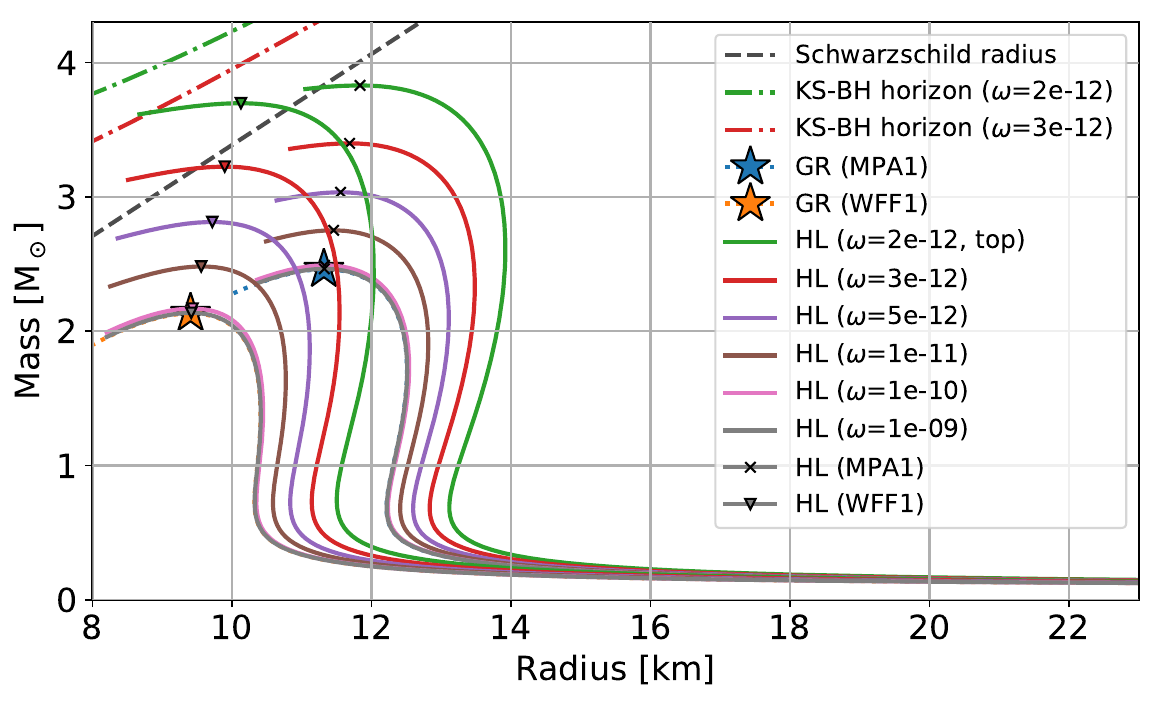}}
  \caption{\label{fig:mrplot} %(Color online)
  Plots of mass-radius relation in \ac{gr} and \ac{hl} with varying parameter $\omega$ for the selected \ac{EOS} models.
  %: \ac{apr4}, \ac{ms1}, \ac{mpa1}, and \ac{wff1}.
  } 
\end{figure}
%%  Mass-Radius plots: end  %%

In Fig.~\ref{fig:mrplot}, we present the mass-radius relations in both \ac{hl} gravity and \ac{gr}, and the relation in \ac{gr} is depicted with the star-marked maximum mass. The heaviest \ac{ns} obtained in \ac{hl} gravity is also marked by different symbols for each \ac{EOS} model. We see that the mass and radius of a \ac{ns} increase as $\omega$ becomes smaller. This result implies that \ac{hl} gravity with a smaller $\omega$ deviates more than \ac{gr}. We also place the horizon of a \ac{sbh} and two horizons of a \ac{ksbh} with $\omega=2\times10^{-12} \mathrm{cm}^{-12}$ and $\omega=3\times10^{-12} \mathrm{cm}^{-12}$ at the top-left corner of Fig.~\ref{fig:mrplot}; we observe that the radii of the maximum stable \acp{ns} in both \ac{hl} gravity and GR under consideration are safely larger than the horizon radius of a black hole with the same mass.
 
%%  omega effect plots: start  %%
\begin{figure}[tbp]
  \centering
       \includegraphics[width=\columnwidth]{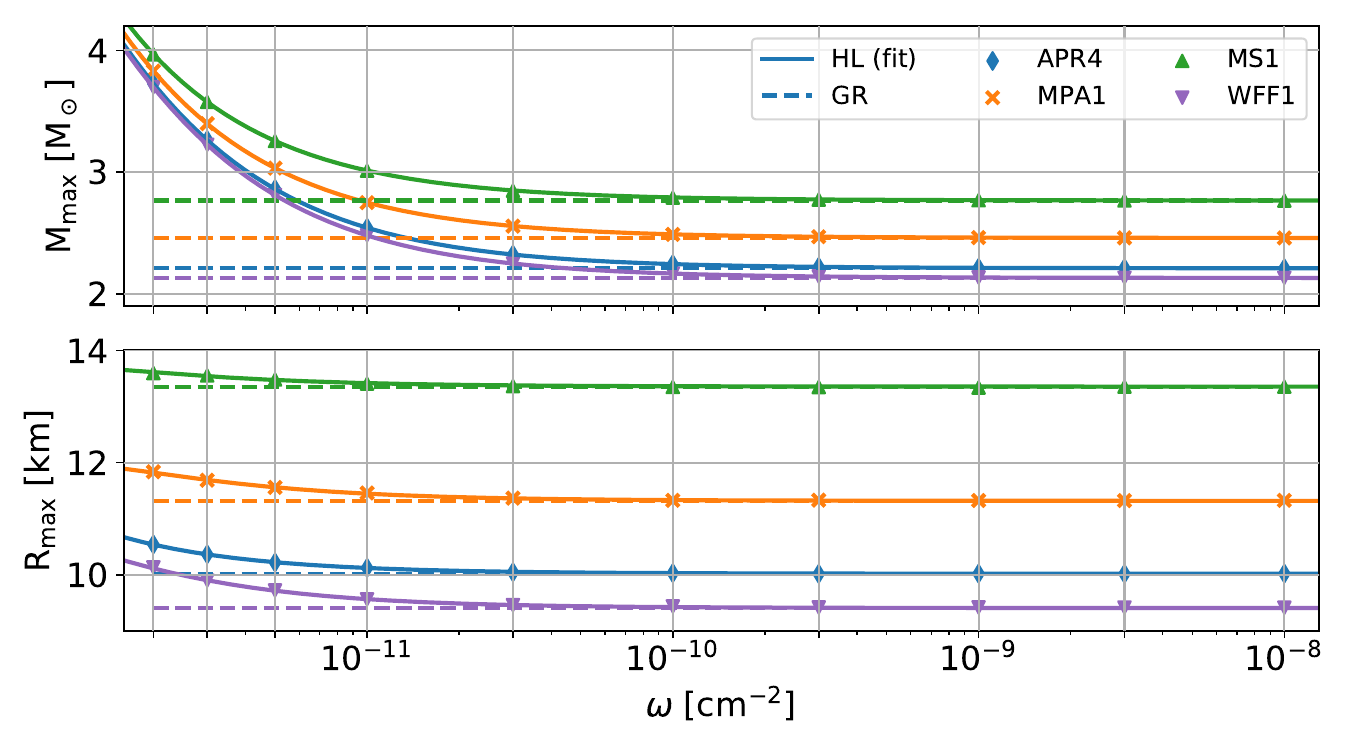}
  \caption{\label{fig:omega} %(Color online)
  Plots of the effect of $\omega$ on the maximum mass and radius for the selected \ac{EOS} models in \ac{gr} and \ac{hl} gravity.}
\end{figure}
%%  omega effect plots: end  %%

We now focus on the $\omega$ dependency in the mass and radius. In Fig.~\ref{fig:omega}, we present the $\omega$-dependent maximum mass (top panel) and radius (lower panel) obtained in \ac{hl} gravity (solid lines) and \ac{gr} (dashed lines) from each of the \ac{EOS} models. We fit the data point with a curve via $M_\text{max}^\text{\acs{hl}} \approx M_\text{max}^\text{\acs{gr}} \left[ 1 + \left( A_1 \omega^{-1} + A_2 \omega^{-2} \right) \right]$ and $R_\text{max}^\text{\acs{hl}} \approx R_\text{max}^\text{\acs{gr}} \left[ 1 + \left( B_1 \omega^{-1} + B_2 \omega^{-2} \right) \right]$ to facilitate the $\omega$ dependency in the mass and radius in \ac{hl} gravity and, consequently, to discuss the $\omega$-dependent difference between \ac{hl} gravity and \ac{gr}. The details of determined fitting coefficients are summarized in Table~\ref{tab1}.
\begin{table}[btp]
\begin{center}
\caption{Fitting coefficients for $M_\text{max}^\text{\acs{hl}}$ and $R_\text{max}^\text{\acs{hl}}$ with all considered \ac{EOS} models.}
\label{tab1}
\centering
\begin{ruledtabular}
\begin{tabular}{ccccc}
\multirow{3}{3em}{\centering \ac{EOS} Models}  & \multicolumn{4}{c}{Fitting Coefficients} \tabularnewline \cline{2-5}
            & \multicolumn{1}{c}{$A_1$}  & \multicolumn{1}{c}{$A_2$} & \multicolumn{1}{c}{$B_1$} & \multicolumn{1}{c}{$B_2$} \tabularnewline 
            & \multicolumn{1}{c}{$(\times \orderAoneB)$}  & \multicolumn{1}{c}{$(\times \orderAtwoB)$} & \multicolumn{1}{c}{$(\times \orderBoneB)$} & \multicolumn{1}{c}{$(\times \orderBtwoB)$} \tabularnewline 
\hline
\ac{apr4}   & \aprAoneB & \aprAtwoB & \aprBoneB & \aprBtwoB \tabularnewline
\ac{mpa1}   & \mpaAoneB & \mpaAtwoB & \mpaBoneB & \mpaBtwoB \tabularnewline
\ac{ms1}    & \msAoneB & \msAtwoB & \msBoneB & \msBtwoB \tabularnewline
\ac{wff1}   & \wffAoneB & \wffAtwoB & \wffBoneB & \wffBtwoB \tabularnewline
\end{tabular}
\end{ruledtabular}
\end{center}
\end{table}

Here, we can see that \ac{hl} gravity can reproduce \ac{gr} when $\omega\gtrsim\omegaGR\textrm{cm}^{-2}$ or $\wGR\ \textrm{cm}^{-2}$ up to tolerances of $|\Delta M_\text{max}| \left( M_\text{max}^\text{\acs{gr}} \right)^{-1}\lesssim \ratioGR$ and $|\Delta R_\text{max}| \left( R_\text{max}^\text{\acs{gr}} \right)^{-1} \lesssim \rGR$, respectively, where $\Delta$ represents the difference between a quantity in \ac{hl} gravity and that in \ac{gr}. We observe that the profile of $M_\mathrm{max}$ and $R_\mathrm{max}$ with $\omega \gtrsim \omegaGR \ {\rm {cm^{-2}}}$ in \ac{hl} gravity has the same behavior as that in \ac{gr} for all the considered \ac{EOS} models. We also observe that the smaller value of $\omega$ in \ac{hl} gravity leads to a larger deviation from \ac{gr}.

\begin{figure}[tbp]
  \centering
  \subfloat[\label{fig:inside_a} mass and gravitational acceleration]{\includegraphics[width=\columnwidth]{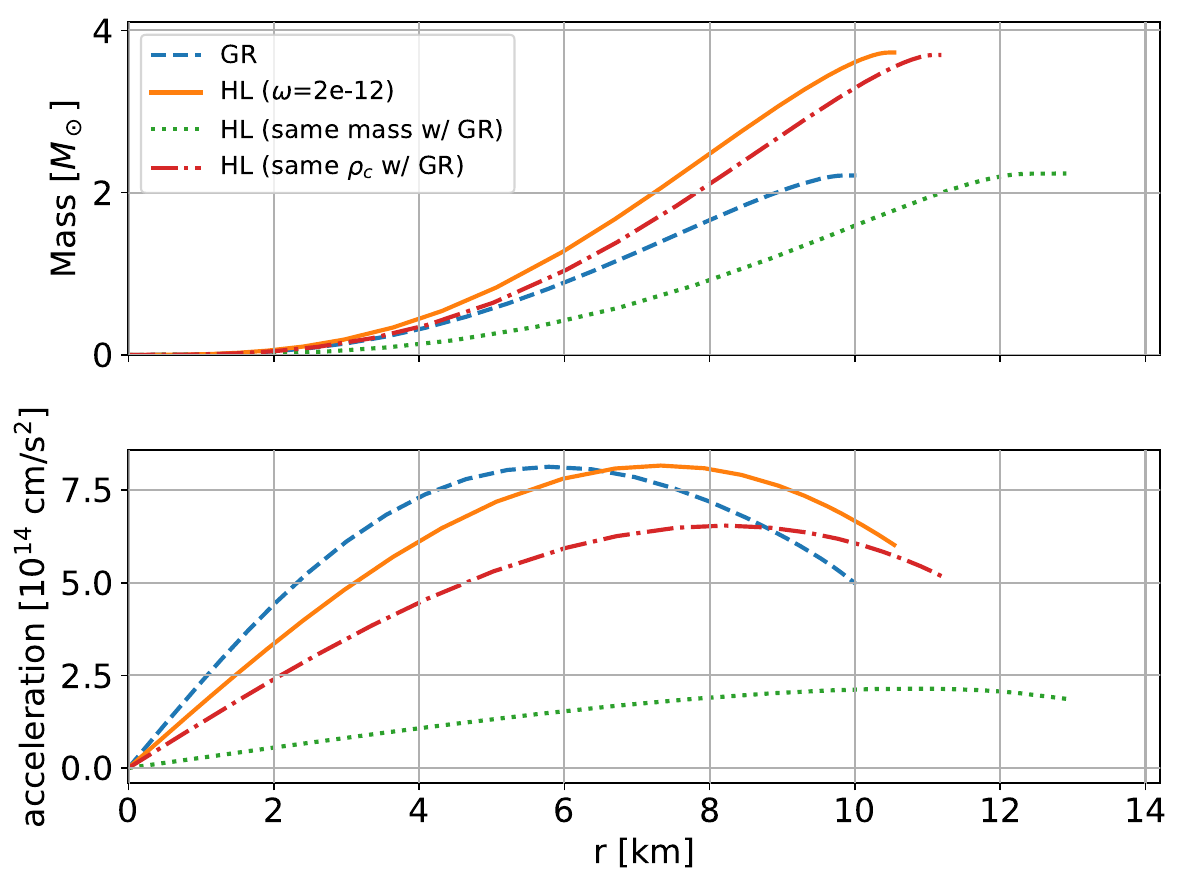}}\\
  \ \ \subfloat[\label{fig:inside_b} density and pressure]{\includegraphics[width=0.97\columnwidth]{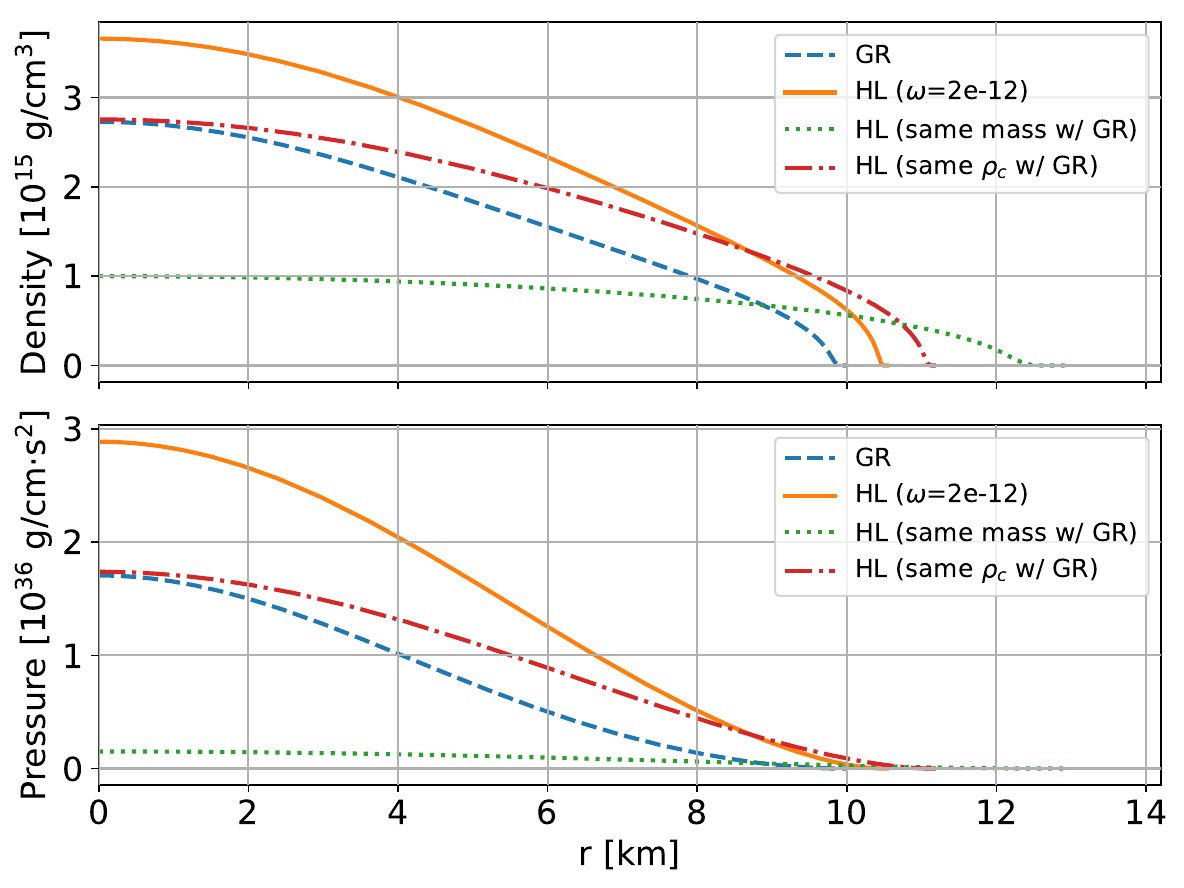}}
  \caption{\label{fig:inside} %(Color online)
  $r$-dependent behaviors of 
  (a) mass and acceleration
  and (b) density and pressure 
  inside the \acp{ns} for the case of \ac{apr4} in \ac{gr} and \ac{hl}.} 
\end{figure}

Next, we examine the $r$-dependent profiles at the interior of the \ac{ns}. We discuss the $r$ dependency with the \ac{apr4} case only because it gives the closest maximum mass in \ac{gr} to the upper limit of the \ac{ns} mass~\cite{TheLIGOScientific:2017qsa} among the selected \ac{EOS} models. In the upper panel of Fig.~\ref{fig:inside} (a), we plot mass profiles of the following \acp{ns}:
\begin{quote}
\begin{tabularx}{0.8\linewidth}{rX}
(i) & The heaviest \ac{ns} in \ac{gr}\\
(ii) & The heaviest \ac{ns} in \ac{hl} gravity\\
{} & ($\omega=2\times10^{-12}\ \textrm{cm}^{-2}$)\\
(iii) & The \ac{ns} in \ac{hl} gravity of the same mass $M^{\text{\acs{gr}}}_{\text{max}}$ as the \ac{ns} in (i)\\
(iv) & The \ac{ns} in \ac{hl} gravity of the same central density $\rho_c$ as the \ac{ns} in (i)
\end{tabularx}
\end{quote}
On the other hand, the gravitational accelerations inside the \acp{ns} corresponding to (i)--(iv) are plotted in the lower panel of Fig.~\ref{fig:inside} (a). The surface gravity of \iac{ns} is given by $g_\text{\ac{ns}}$~\cite{Shapiro:1983du,Bejger:2004gz}. Similarly, the gravitational acceleration experienced by an observer on a fixed $r$ inside \iac{ns} can be calculated as $a(r) = (1 - f(r))^{1/2} \Phi'(r)$, which naturally becomes $g_\text{\ac{ns}}$ at the surface $R$. We observe that \ac{hl} gravity has a relatively weaker gravitational acceleration than that in \ac{gr} near the central region of the \ac{ns}. Even though the \ac{ns} in (ii) is much heavier than the \ac{ns} in (i), its gravity is weaker at $r \lesssim 6\ \mathrm{km}$. This observation explains why the heavier \ac{ns} rather than the \ac{ns} in (i) remains stable in \ac{hl} gravity without collapsing into a black hole. The weaker gravity of \ac{hl} is also observed from the curve of the same-mass \ac{ns} in (iii) at the bottom panel of Fig.~\ref{fig:inside} (a). Note that the maximum accelerations inside the heaviest
\acp{ns} are almost the same in both \ac{gr} and \ac{hl} gravity; however, the accelerations
may be slightly different depending on the \ac{EOS} models.

%%  inside NS plots: start  %%
%%  inside NS plots: end  %%

In addition, the profiles of the density and pressure inside \ac{ns}s are also shown in the upper panel and lower panel, respectively, in Fig.~\ref{fig:inside} (b). Due to the weaker gravitational attraction inside the same-mass \ac{ns} in (iii), the pressure and density increase slowly as $r$ goes to zero, and the central density becomes much smaller than that of the \ac{ns} in (i), which is consistent with the larger radius of the \ac{ns} in (iii). The central density of the \ac{ns} in (iv) is the same as that of the \ac{ns} in (i), and the density and pressure decrease slowly as $r$ increases and vanish at larger radii, with the result that the \ac{ns} in (iv) becomes naturally heavier than the \ac{ns} in (i). Note that the weaker gravitational attraction in \ac{hl} gravity can be easily understood by computing the Kretschmann curvature scalar defined by $\kret \equiv {\mathcal R}_{\alpha\beta\gamma\sigma}{\mathcal R}^{\alpha\beta\gamma\sigma}$. Indeed, the Kretschmann curvature scalar of the \ac{ksbh} solution can be compared to that of the \ac{sbh} solution with the same mass and radius. In \ac{gr}, $\kret_\text{\acs{sbh}} = 48 G_N^2 c^{-4} M^2 r^{-6}$, whereas $\kret_\text{\acs{ks}BH} \xrightarrow{r \ll r_c} (81 / 4) G_N c^{-2} M \omega r^{-3} [ 1 + O(r / r_c)^{3/2} ]$ and $\kret_\text{\acs{ks}BH} \xrightarrow{r \gg r_c} \kret_\text{\acs{sbh}} [ 1 + O(r_c / r)^{3} ]$ in \ac{hl} gravity, where $r_c \equiv (4 G_N c^{-2} M \omega^{-1})^{1/3}$ is a characteristic scale of the \ac{ksbh} solution. The overall behavior of the Kretschmann curvature scalar for the \ac{ksbh} solution in \ac{hl} is slowly growing, unlike the \ac{sbh} solution in \ac{gr}, which implies that the gravitational force induced by the curvature scalar in \ac{hl} gravity is weaker than that in \ac{gr} in the limit of $r\rightarrow 0$. Then the mass obtained by integrating Eq.~(\ref{Eq_tov_mass}) from $r=0$ to $r=R$ also becomes heavier, eventually. However, the results of $M_{\textrm{max}} \gtrsim 3 \msun$ in \ac{hl} gravity are placed not only beyond the upper limit of the theoretical estimation of the upper limit~\cite{ULnsTh:1996aap,Kalogera:1996ci} but far beyond the upper limit of the mass of a \ac{ns}~\cite{2solar:2010nature,2solar:2018apj,Margalit:2017dij,ULns:2017prd,ULns:2018prd,ULns:2018apjl}.

%%%%%%%%%%%%%%%%
%%  Discussion  %% 
%%%%%%%%%%%%%%%%
\section{Discussions}
\label{sec:discussion}

We have discussed the characteristics of \ac{hl} gravity by examining the $\omega$ dependency of the physical observables, especially, mass and radius, of a \ac{ns}. The considered range of $\omega \ge 2 \times 10^{-12} \ \text{cm}^{-2}$ has been chosen based on the lower bounds determined by observations, $\omega_\text{min} \sim 10^{-16} \ \text{cm}^{-2}$, and the cosmic censorship for a $4\ \msun$ black hole candidate, GRO J0422+32, that reads $\omega \ge 1.433 \times 10^{-12} \ \textrm{cm}^{-2}$.

The conventional measurement of the mass and radius of a \ac{ns} is conducted by observations on a pulsar, rotating \ac{ns}, consisting of a binary system. Depending upon the type of pulsar, different approaches have been implemented to measure or to constrain the mass and radius (for more details, see Ref.~\cite{Ozel:2016araa} and the references therein). In addition, from the recent detection of GW170817, radiated from a merger of binary \ac{ns} systems, it is now  possible to consider the \ac{gw} as an additional approach to estimating the mass and radius of a \ac{ns}~\cite{TheLIGOScientific:2017qsa,Abbott:2018exr}. In addition, a possible problem of the exclusion of \ac{hl} gravity was raised with the help of GW170817 by a comparison between the speeds of \ac{gw} and the electromagnetic wave, the polarization of \acp{gw} and so on~\cite{Creminelli:2017sry, Ezquiaga:2017ekz, Yagi:2013ava, Yagi:2013qpa}. However, there are also unconstrained parameter regions that favor \ac{hl} gravity even after the observation of GW170817~\cite{Barausse:2019yuk}, and the problem requires further studies. Hence, to discuss the viability of \ac{hl} gravity with the physical observables of a \ac{ns}, it is worthwhile to address some aspects of \ac{hl} gravity more rigorously such as (i) whether it can pass the tests on post-Keplerian parameters~\cite{Ozel:2016araa}, (ii) whether it is possible to derive \ac{gw} waveforms with \ac{hl} gravity to perform parameter estimation as done with \ac{gr}, and (iii) how the tidal deformability of a \ac{ns} in \ac{hl} gravity differs from that in \ac{gr}, and so on.\footnote{Our analysis is only for the nonprojectable case of \ac{hl} gravity, which yields a stable static solution of a \ac{ns}. For the projectable case, see Ref.~\cite{Izumi:2009ry}, there is no stable static solution.}

\ac{hl} gravity is also open to the possibility of being verified through other compact objects such as black holes and white dwarfs. \ac{hl} gravity was originally designed for recovering the usual asymptotic property of \ac{gr} whereas it has a UV modification in the strong gravity region unlike \ac{gr}. An observer in the IR region can always experience the asymptotic structure of the object rather than the UV structure. The feature of the UV structure is only reflected in the physical parameters such as mass, spin, and so on. Thus, one finds that the stellar mass black hole behaves consistently no matter how large/small its mass is once the black hole forms whereas the only differences are the deviation of physical parameters encoded by the UV structures of black holes. As shown in the \ac{ks} solution, it clearly looks like a \ac{sbh} at large distances of $r$ whereas it differs from the \ac{sbh} as $r$ goes to the UV region. The extent of the difference between \ac{hl} gravity and \ac{gr} depends upon the parameter $\omega$. The relatively large $\omega$ leads to the small deviation from \ac{gr} because we have the form of the \ac{ks} solution: $f(r) = 2 k / r (1  - k/ \omega r^3 +\cdots)$, where $k= G_N M/c^2$. Therefore, we may constrain the lower bound of $\omega$ further from the mass of the stellar mass black holes which can be estimated by the \ac{gw} and/or x-ray observations.

%%  Mass-Radius for WD plots: start  %%
\begin{figure}[tbp]
  \centering
  \includegraphics[width=\columnwidth]{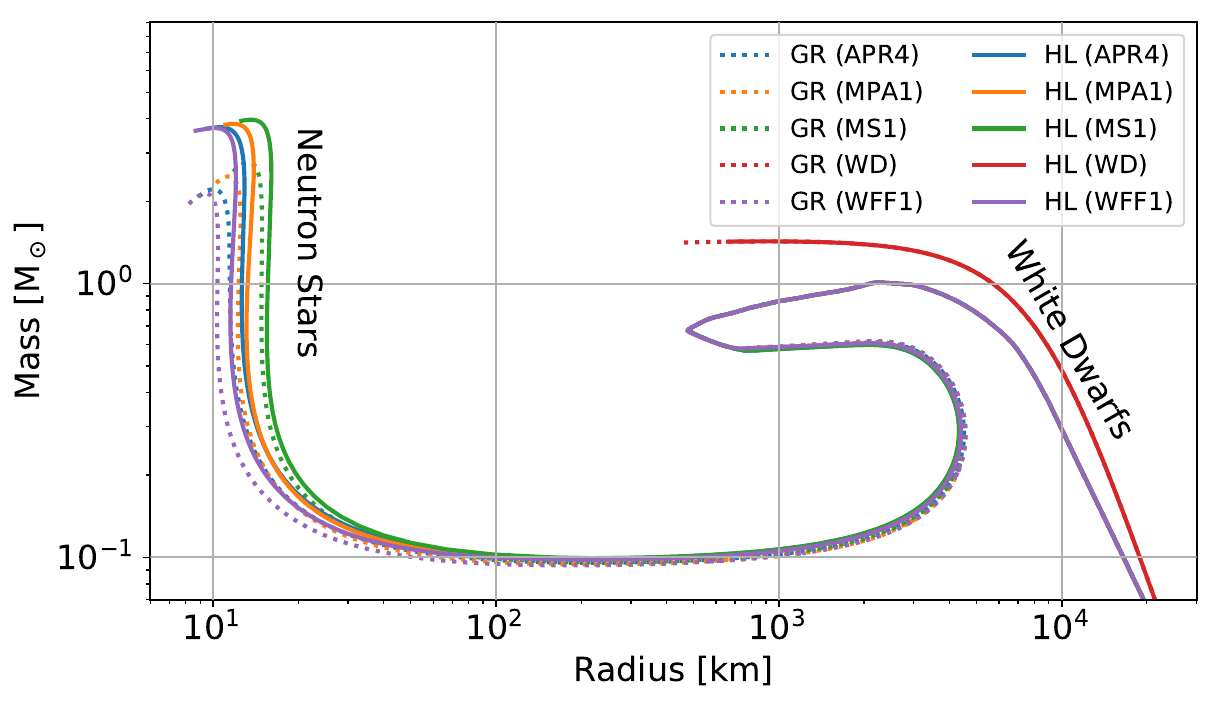}
  \caption{\label{fig:whitedwarf} %(Color online) 
  Plots of mass-radius relation in \ac{gr} and \ac{hl} with $\omega = 2 \times 10^{-12} \ \mathrm{cm}^{-2}$ for the \ac{EOS} models for \acp{ns} under consideration (\ac{apr4}, \ac{ms1}, \ac{mpa1}, and \ac{wff1}) and the relativistic \ac{EOS} for white dwarf (WD).}
\end{figure}
%%  Mass-Radius for WD plots: end  %%

Now, it is noteworthy that the \ac{EOS} models for the \ac{ns} structure do not show significant differences in the structure of a white dwarf. This is primarily because they are almost identical except for the physics at the nucleus size. Similarly, \ac{hl} gravity deforms the behavior at \ac{uv} scale and does not change the structure of a white dwarf significantly. The behaviors of white dwarfs as well as \acp{ns} in \ac{gr} and \ac{hl} gravity are depicted in Fig.~\ref{fig:whitedwarf}. The relativistic \ac{EOS} for white dwarfs is obtained by assuming that the dominant chemical components of white dwarfs are $^4$He, $^{12}$C, or $^{16}$O, that is, the ratio between the atomic mass number $A$ and the number of electrons $Z$ is given by $A / Z = 2$~\cite{Sagert:2005fw}. The \ac{EOS} models for \acp{ns} do not reflect the Chandrasekhar limit correctly, because they are obtained by considering physical situations beyond the degeneracy pressure of electrons such as neutronization. The mass-radius relation of white dwarfs in the {\it beyond Horndeski class of gravity theories} was investigated in~\cite{Saltas:2018mxc}.

In conclusion, it is shown that \ac{hl} gravity reveals a deviant feature from GR in the short distance regime due to its relatively weaker gravitational force compared with that of \ac{gr}. We observe that the deviation is sensitive to the choice of parameter $\omega$. In addition, the \ac{ns} with the larger radius and the heavier mass is far beyond the upper limit of the current observational results obtained from \ac{gr} or post-Keplerian. To validate \ac{hl} gravity, theoretical investigations and future observations of \acp{gw} from compact binary mergers containing at least one \ac{ns} will yield more constraints on the physical observables of a \ac{ns} and, eventually, will determine the fate of \ac{hl} gravity.

%%%%%%%%%%%%%%%%%%%%%%%
%%  Acknowledgments  %%
%%%%%%%%%%%%%%%%%%%%%%%
\acknowledgments 
J.J.O. would like to thank Parada Hutauruk for helpful discussions at the early stages of this work. The authors would like to thank Mu-In Park for fruitful discussions on \ac{hl} gravity,  Chang-Hwan Lee and Young-Min Kim for useful discussions on neutron stars and tidal deformations related to GW170817, and Olivier Minazzoli, Nathan Johnson-McDaniel, Noah Sennett, Harald Pfeiffer, Tjonnie Li, and Chris Van Den Broeck for helpful comments and discussion on this work. This work was supported by Basic Science Research Program through the National Research Foundation of Korea (NRF) funded by the Ministry of Education (Grants No. NRF-2020R1I1A2054376 and No. NRF-2018R1D1A1B07041004), the 
NRF grant funded by the Ministry of Science and ICT (Grant No. NRF-2020R1C1C1005863), and the National Institute for Mathematical Sciences funded by the Ministry of Science and ICT (Grant No. B20710000). K.K. was also supported by grants from the Research Grants Council of Hong Kong (Projects No. CUHK 14310816 and No. CUHK 24304317) and the Direct Grant for Research from the Research Committee of the Chinese University of Hong Kong.
\bibliographystyle{apsrev4-1}
\bibliography{TOV,HL,gw,RK}

\end{document}